\documentclass[10pt,conference]{IEEEtran}
\IEEEoverridecommandlockouts
\usepackage{cite}
\usepackage{amsmath,amssymb,amsfonts}
\usepackage{algorithmic}
\usepackage{graphicx}
\usepackage{textcomp}
\usepackage{xcolor}
\newcommand{\alg}{\textsc{RADICE}}
\usepackage{comment}
\usepackage{paralist}
\usepackage[inline]{enumitem}
\usepackage[ruled,noend,linesnumbered]{algorithm2e}
\usepackage{multirow}
\usepackage{rotating}
\usepackage[hidelinks]{hyperref}
\usepackage{booktabs}
\usepackage{tabularx}
\usepackage{subcaption}
\usepackage{balance}
\usepackage{mwe}
\usepackage{tikz}

\newcommand\copyrighttext{%
  \footnotesize \textcopyright \the\year{} IEEE. Personal use of this material is permitted.  Permission from IEEE must be obtained for all other uses, in any current or future media, including reprinting/republishing this material for advertising or promotional purposes, creating new collective works, for resale or redistribution to servers or lists, or reuse of any copyrighted component of this work in other works.}
\newcommand\copyrightnotice{%
\begin{tikzpicture}[remember picture,overlay]
\node[anchor=south,yshift=10pt] at (current page.south) {\fbox{\parbox{\dimexpr\textwidth-\fboxsep-\fboxrule\relax}{\copyrighttext}}};
\end{tikzpicture}%
}

\def\BibTeX{{\rm B\kern-.05em{\sc i\kern-.025em b}\kern-.08em
    T\kern-.1667em\lower.7ex\hbox{E}\kern-.125emX}}
\begin{document}

\title{\alg: Causal Graph Based Root Cause Analysis for System Performance Diagnostic
\thanks{$^{\dag}$Corresponding author (andrea.tonon1@huawei-partners.com).}
}

\author{\IEEEauthorblockN{Andrea Tonon,$^{1,\dag}$ Meng Zhang,$^{2}$ Bora Caglayan,$^1$ Fei Shen,$^2$ Tong Gui,$^2$ MingXue Wang,$^1$ Rong Zhou$^2$ }
\IEEEauthorblockA{\textit{$^1$Huawei Ireland Research Center}, Dublin, Ireland \\
\textit{$^2$Huawei Nanjing R\&D Center}, Nanjing, China}}

\maketitle

\copyrightnotice

\emph{Radice}: (Italian noun) root.
\newline
\begin{abstract}
Root cause analysis is one of the most crucial operations in software reliability regarding system performance diagnostic. It aims to identify the root causes of system performance anomalies, allowing the resolution or the future prevention of issues that can cause millions of dollars in losses. Common existing approaches relying on data correlation or full domain expert knowledge are inaccurate or infeasible in most industrial cases, since correlation does not imply causation, and domain experts may not have full knowledge of complex and real-time systems.
In this work, we define a novel causal domain knowledge model representing causal relations about the underlying system components to allow domain experts to contribute partial domain knowledge for root cause analysis. We then introduce \alg, an algorithm that through the causal graph discovery, enhancement, refinement, and subtraction processes is able to output a root cause causal sub-graph showing the causal relations between the system components affected by the anomaly. We evaluated \alg\ with simulated data and reported a real data use case, sharing the lessons we learned. The experiments show that \alg\ provides better results than other baseline methods, including causal discovery algorithms and correlation based approaches for root cause analysis.
\end{abstract}

\begin{IEEEkeywords}
Root cause analysis, System performance diagnostic, Causal discovery, Causal domain knowledge modeling
\end{IEEEkeywords}

\section{Introduction}
Root cause analysis is one of the most crucial operations in software reliability regarding system performance diagnostic. Given a time series representing a system performance metric that exhibits an anomalous behavior, the goal of root cause analysis for system performance diagnostic is to detect which components of the considered system caused the anomaly that degraded its performance.
In particular, an anomaly is an event that changes the distribution of a system component metric and then propagates to all other system components causally related to it over time, up to the system performance metric.
In the context of system performance diagnostic, examples of system component metrics are CPU/memory usage of different components, while system latency or connection time are possible system performance metrics.  
A drop in one of these performance metrics indicates the presence of an issue in the system, which can cause millions of dollars in losses. Thus, being able to accurately detect which of the system components caused the anomaly by taking into account the system causal structure is a problem of crucial importance in solving or preventing similar future issues.
\begin{figure}[]
 \centerline{\includegraphics[width=\linewidth]{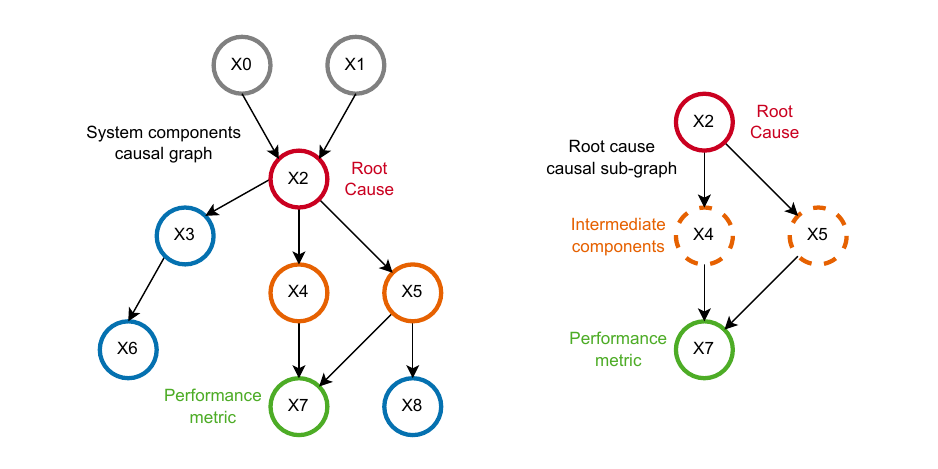}}
	\caption{Example of a whole causal graph of a system (left) and of a root cause causal sub-graph associated with an anomaly in the same system (right). Nodes represent system component metrics and performance metrics, and edges represent causal
relations between them. $X7$ is the performance metric of the system. $X2$ is the component that caused the anomaly (the root cause of the anomaly). $X3$, $X4$, $X5$, $X6$, $X7$, and $X8$ are all components affected by the anomaly: while $X4$ and $X5$ are intermediate components between the root cause and the performance metric, $X3$, $X6$, and $X8$ do not have a causal impact on the performance metric.}
	\label{fig:exCau}
\end{figure} 

In such a scenario, current approaches have several limitations.
Commonly used root cause analysis methods are correlation based algorithms~\cite{shan2019diagnosis,luo2014correlating} that in many scenarios may not be reliable since correlation does not imply causation~\cite{pearl2009causality}. Thus, they are unable to discriminate between causal relations and spurious associations. For example, in Fig.~\ref{fig:exCau}-left, both $X7$ and $X8$ are caused by $X5$ and thus they can be highly correlated, but they are not causally related.
To provide more reliable results, many causal based root cause analysis algorithms require the knowledge of a complete causal graph about the underlying system defined by domain experts~\cite{gan2021sage,qiu2020causality}, like the one shown in Fig.~\ref{fig:exCau}-left. However, in many real industrial cases, it is not realistic to have a complete and accurate structural causal model of the underlying system since the data may flow through different components controlled and developed by different teams. It is also not flexible, since metrics can frequently change based on the system evolution.
To avoid the manual definition of accurate causal graphs, some root cause analysis methods employ causal discovery algorithms to directly learn from monitoring data the causal relations between the system components. However, causal discovery methods provide guarantees to recover the correct causal graph only under data distribution assumptions~\cite{guo2020survey,pearl1995causal} that may be difficult to satisfy and verify in real-world scenarios, leading to the discovery of spurious and unoriented causal relations and the absence of true causal relations.
Finally, causal based root cause analysis algorithms often aim to only identify the set of root causes, without reporting the causal sub-graph that relates them~\cite{ikram2022root}, like the one shown in Fig.~\ref{fig:exCau}-right. Thus, identifying the causal sub-graph that shows how the anomalies have spread through the system is another key challenge.
To the best of our knowledge, no existing approach provides solutions for all these limitations.

To address all the above challenges, in this work we introduce \alg, a causal graph based root cause analysis algorithm for system performance diagnostic. \alg, by enhancing a causal discovery method~\cite{guo2020survey,runge2020discovering}, is able to automatically learn the causal relations between the system components from the input monitoring data. In particular, by employing an entropy based orientation strategy and by allowing a (partial) causal domain knowledge model in input, it can overcome the above mentioned limitations.
The idea is to allow domain experts to contribute partial domain knowledge to enhance the causal discovery phase and to employ an entropy based orientation strategy to orient causal relations that causal discovery algorithms are not able to orient, in order to learn a more robust causal graph representing the causal relations between the system components. 
Hence, manually defining a complete causal graph or totally relying on causal discovery algorithms is not required.
The partial domain knowledge is also employed by \alg\ in the causal graph refinement phase, in combination with an adjusted correlation score, to identify components that cannot be root causes of the anomaly, avoiding further spurious associations.
Finally, it trims the causal graph considering the remained candidate root causes, obtaining the root cause causal sub-graph that represents the causal relations between the performance metric and the root causes. The root cause causal sub-graph and a time series plot of the related metrics are the outputs of \alg.

\begin{figure*}[htbp]
 \centerline{\includegraphics[width=\textwidth]{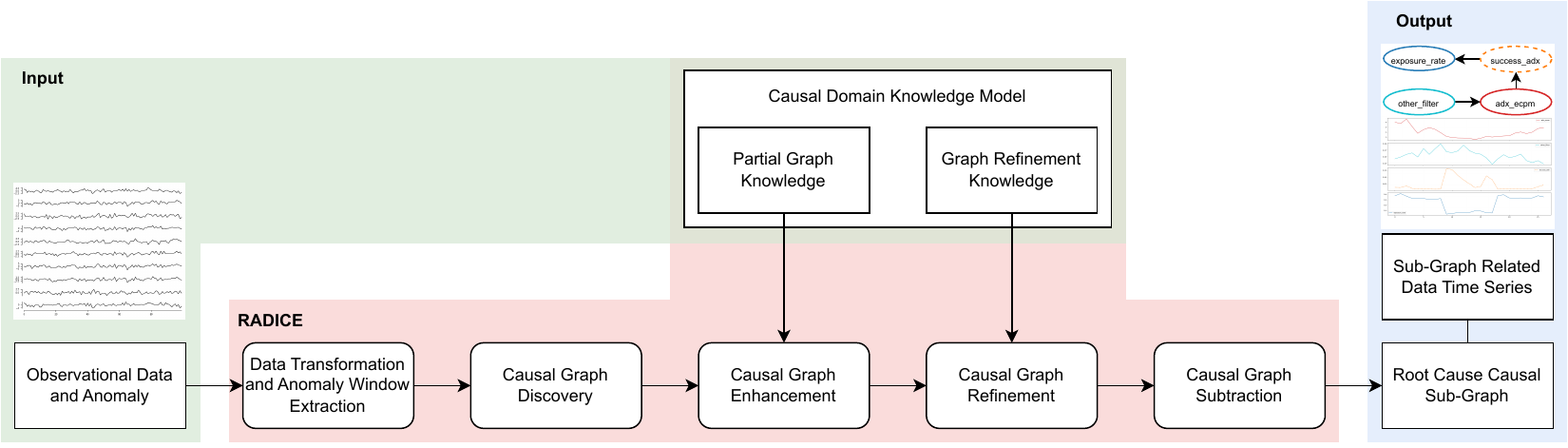}}
	\caption{Schema of \alg.}
	\label{fig:algo}
\end{figure*}

In this regard, our contributions are as follows:
\begin{itemize}
    \item We introduce \alg, our method that through the causal graph discovery, enhancement, refinement, and subtraction processes, outputs a root cause causal sub-graph showing the causal relations between system components causally related to the anomaly. To enhance the root cause analysis, \alg\ allows (partial) domain knowledge in input. (See Fig.~\ref{fig:algo} for a detailed schema.)
    \item We introduce a causal domain knowledge model to allow domain experts to contribute partial causal knowledge for a causal based root cause analysis method, such as \alg, since in many industrial scenarios it is not realistic to have a full causal model of complex systems.
    \item For the graph enhancement process, we introduce a strategy to enhance a causal graph obtained by a causal discovery algorithm employing our causal domain knowledge model and an entropy based orientation strategy to correct spurious, unoriented, and missing causal relations.
    \item For the graph refinement process, we describe a strategy to identify system components that cannot be root causes by combining an adjusted correlation score with our causal domain knowledge model. 
    \item We perform an experimental evaluation with simulated and real system data demonstrating that \alg\ outperforms other baseline methods and that it can find root cause causal sub-graphs in real systems.
\end{itemize}

\section{Related Works}
We now discuss the relation of our work to prior art on root cause analysis, causal discovery, and causal based root cause analysis.
Root cause analysis is widely used to ensure the reliability of production systems~\cite{lin2020fast,chen2019outage} in many domains such as IT operations~\cite{soldani2022anomaly}, telecommunications, etc. Commonly used root cause analysis algorithms are based on correlation~\cite{dynatrace,shan2019diagnosis,su2019coflux}. Thus, they may not be reliable in complex industrial scenarios, such as system performance diagnostic, since correlation does not imply causation. For example, \cite{su2019coflux} introduced a strategy to compute features that highlight data anomalies. However, it considers the correlation between the features to identify the root causes, and thus it is not able to discriminate between spurious and causal relations.
Even if \alg\ employs a strategy similar to previous works in its refinement phase, we consider both causal relations and an adjusted correlation score~\cite{dynatrace}, with the latter employed only to identify components that cannot be root causes, eventually in combination with domain knowledge.

Causal discovery is the task of learning causal graphs representing causal relations between entities from observational data. To solve such a problem, several algorithms have been proposed~\cite{pearl1995causal,runge2020discovering,spirtes2000causation}, including approaches based on neural networks~\cite{nauta2019causal}. (We refer the readers to a survey~\cite{guo2020survey} for a thorough discussion.) However, it is well known that any causal conclusion drawn from observational data requires assumptions that are not testable in the observational environment~\cite{guo2020survey,pearl2009causal}. Thus, in this work, we introduce a method to enhance the causal graph obtained with a causal discovery algorithm using partial domain knowledge defined by domain experts and an entropy based orientation strategy~\cite{kocaoglu2017entropic}. Such a method is employed by \alg\ to obtain a more robust causal graph that can be used for root cause analysis.

More recently, root cause analysis algorithms have started to employ causal discovery algorithms and/or domain knowledge to consider causal relations between metrics~\cite{budhathoki2022causal,ikram2022root,li2022causal,qiu2020causality}, with multiple works~\cite{6848128,9678708,9527007} aiming to build causal graphs to diagnose complex systems.
In particular, \cite{qiu2020causality} utilizes domain knowledge to improve the performance of the PC algorithm~\cite{pearl1995causal}, but the improvements only affect the computational complexity, while we leverage domain knowledge to obtain a more robust causal graph. In addition, by employing the PC algorithm, it cannot consider lagged relations that are essential in time series data. Finally, it considers as root causes all the components connected with the performance metric, without filtering out possible false associations, as \alg\ does. Instead, \cite{budhathoki2022causal} and~\cite{li2022causal} require full knowledge of the underlying system that is almost always infeasible to have. In contrast, \alg\ allows partial domain knowledge to improve the causal discovery performance and filter out spurious root causes. 
Further works~\cite{ma2020automap,9213058} do not incorporate domain knowledge and rely solely on (modified) causal discovery algorithms, retaining the issues of such methods. Additional causal based root cause analysis methods~\cite{ikram2022root} aim to only identify the root causes and therefore are unable to output causal sub-graphs. Thus, they are not in the scope of this work.

To the best of our knowledge, \alg\ is the first root cause analysis method to employ a causal discovery algorithm, together with a causal domain knowledge model which allows partial knowledge, to output a root cause causal sub-graph, and our evaluation demonstrates its effectiveness.

\section{Causal Graph Based Root Cause Analysis}
\label{sec:method}
In this section, we introduce the overall architecture of our method \alg, causal g\underline{RA}ph base\underline{D} root cause analys\underline{I}s for system performan\underline{CE} diagnostic. Figure~\ref{fig:algo} shows a schema with all its components. The inputs of \alg\ are a \emph{time series dataset} $\mathbf{X} = \{X_1, X_2,\dots,X_N\}$ containing $N$ observed numerical time series $X_i \in \mathbb{R}^T$ of the same length $T \in \mathbb{N}^+$ and a \emph{causal domain knowledge model} $\mathcal{DK}$. 
The time series dataset $\mathbf{X}$ is composed of a \emph{system performance metric} $X_t \in \mathbf{X}$ that exhibits an anomalous behavior and a set of \emph{candidate metrics} $\mathcal{C} = \mathbf{X} \setminus X_t$ that may have caused the anomaly in $X_t$, each representing a system component metric. Instead, the causal domain knowledge model $\mathcal{DK}$, composed of the \emph{partial graph knowledge} and \emph{graph refinement knowledge}, represents causal knowledge, defined by domain experts, about the system components. The causal domain knowledge model is used by \alg\ to correct spurious relations that may be found in the data, to include causal relations that may be missed by the causal discovery algorithm, and to filter out further candidate metrics that cannot be causally related to the anomaly. More details about the causal domain knowledge model are provided in section~\ref{sec:dk}. The aim of \alg\ is to construct a \emph{root cause causal sub-graph} $\mathbf{G_{RC}}$ (see Fig.~\ref{fig:exCau}-right) that represents the causal relations between the system performance metric $X_t$ and the candidate metrics that may have caused the anomaly in $X_t$, i.e., the \emph{root causes} $\mathcal{RC}$. The outputs of \alg\ are then the root cause causal sub-graph $\mathbf{G_{RC}}$ and the time series plot of the metrics involved. 

First, the input dataset $\mathbf{X}$ is pre-processed. In this phase, data transformations, such as data cleaning and rate conversion, are applied to $\mathbf{X}$. (Note that such transformations depend on the input data and on the system considered. For example, to convert metrics representing integer values into rates considering their respective total counts allows to reduce the number of metrics and thus to reduce possible noise in the data.) Then, an anomaly detection algorithm is applied to the transformed data. The goal of the anomaly detection is to extract the time window in which the performance metric $X_t$ exhibits the anomalous behavior, representing a drop in the system performance. Such a time window is then used to select the data that will be used in the following phases. Given $n$ the number of time samples in the anomaly window, in the following phases \alg\ considers the $n$ time samples that precede the anomaly window, the $n$ time samples of the anomaly window, and the $n$ time samples that follow the anomaly window, for all the considered metrics, to focus the causal discovery on anomalous data associated with a single anomaly of the system. Note that in the following sections, for the input dataset $\mathbf{X}$, and thus $X_t$ and the candidate metrics $\mathcal{C}$, we are only considering such a portion of each time series.

The core part of \alg\ is composed of four phases: \emph{causal graph discovery}, \emph{causal graph enhancement}, \emph{causal graph refinement}, and \emph{causal graph subtraction}. 
The causal graph discovery aims to learn the causal relations between all the component metrics by executing a causal discovery algorithm in $\mathbf{X}$. However, causal discovery is a complex task, and causal discovery algorithms may report false positives, report unoriented causal relations, and miss true causal relations, due to noise and uncertainty in the data.
Thus, in the causal graph enhancement, \alg\ leverages the partial graph knowledge and an entropy based orientation strategy to correct some of these problems. 
Then, in the causal graph refinement, \alg\ combines a correlation based approach with the graph refinement knowledge to filter out candidate metrics that cannot be the root causes of the anomaly, obtaining the candidate root causes $\mathcal{RC}$. 
Finally, in the causal graph subtraction, \alg\ trims the causal graph to obtain the root cause causal sub-graph $\mathbf{G_{RC}}$ which represents the causal relations between $X_t$ and the root causes $\mathcal{RC}$ that have a causal path with it and are causally related to the anomaly.

\subsection{Causal Domain Knowledge Model}
\label{sec:dk}
In this section, we describe the \emph{causal domain knowledge model} used as input by \alg.
The causal domain knowledge model allows domain experts to contribute partial knowledge to the causal graph used in the root cause analysis process, since some causal relations may not be directly deductible from the data. Thus, \alg\ differs from many root cause analysis frameworks that require a complete and accurate structural causal model provided by domain experts. We believe that it is not realistic to provide a complete and accurate structural causal model in complex business scenarios since the data may flow through different sub-systems controlled and developed by different teams. It is also not flexible, since components can be frequently added and removed with the system evolution.
In particular, \alg\ considers a causal domain knowledge model defined by domain experts to correct spurious relations that may be found in the data and to include causal relations that may be missed by the algorithm. 
The causal domain knowledge model includes the partial graph knowledge and the graph refinement knowledge.

\subsubsection{Partial Graph Knowledge}
The \emph{partial graph knowledge} represents causal relations, defined by domain experts, between the involved system components. In this work, we represent it as a set of \emph{domain directed edges} $E_D$ between pairs of component metrics and a \emph{domain node level function}. 
The domain directed edges $E_D$ represent directed and instantaneous causal relations between pairs of component metrics, i.e., if there is a directed edge $(u,v) \in E_D$ starting from metric $u$ to metric $v$, then $u$ causes $v$ without time-lag. Thus, even if the domain experts do not have full knowledge of the causal graph, like the one shown in Fig.~\ref{fig:exCau}-left, they often know some causal relations between given pairs of metrics, like some of the single edges shown in Fig.~\ref{fig:exCau}-left, that can be added to a causal graph learned from monitoring data whether the causal discovery algorithm missed them (see sections~\ref{sec:cauDis} and~\ref{sec:cgdw}). 
(Note that the domain directed edges must not create cycles to represent valid instantaneous causal relations.) 
Instead, the domain node level function, denoted by $\texttt{level}(\cdot)$, represents higher-order causal relations between the component metrics. It associates each metric with a level that represents its higher-order causal relations with the other metrics. For example, the levels can be defined considering the topological structure of the targeted system, assigning to the performance metric the highest level. Then, a metric $u$ cannot be caused by a metric $v$ if $u$ has a lower level than $v$ ($\texttt{level}(u) < \texttt{level}(v)$).
Instead, the metric $v$ can cause metrics of its same level or higher.
Let us consider, as an example, a system composed of 2 components A and B (each of them described by many metrics, such as request count, resource utilization, etc.). If data flows from A to B, then an anomaly in B cannot propagate to A, and thus A's metrics will have a lower level than B's metrics. This allows to define higher-order causal relations between the system component metrics even if the directed causal relations between them are unknown. Considering the causal graph shown in Fig.~\ref{fig:exCau}-left, domain experts may know that $X3$, $X4$, and $X5$ cannot cause $X0$, $X1$, and $X2$, for example for the type of components they represent. Thus, the former would have a higher level with respect to the latter, and this information can be used to orient unoriented causal relations that may be found by the causal discovery algorithm (see sections~\ref{sec:cauDis} and~\ref{sec:cgdw}).
Finally, let us remember that the domain directed edges must not create cycles and that they must be defined in accordance with the higher-order relations of the domain node level function.

\subsubsection{Graph Refinement Knowledge}
\label{sec:dcr}
The \emph{graph refinement knowledge} is a set of rules used by \alg\ to find, between the candidate metrics $\mathcal{C}$, those that are associated with the anomaly, i.e., the set $\mathcal{RC}$. They represent necessary (but not sufficient) conditions for the candidate metrics $\mathcal{C}$ to be root causes. Depending on the data or the system, such rules can be defined in different ways. In this work, we consider the sign that the correlation between the performance metric $X_t$ and a candidate metric $X_i \in \mathcal{C}$ must have to consider $X_i$ a candidate root cause.
Consider, for example, that one of the candidate metrics represents the fail rate of a system component. An increase in its value over time indicates that there are some problems in the system, while a decrease in its value represents an improvement of the system performance. Then, such a metric must have a negative correlation with $X_t$ to be considered a candidate root cause, since a positive correlation would mean that an improvement of the system performance caused the anomaly. 
Thus, when in the input data we observe high adjusted correlation scores (see section~\ref{sec:rcv}) between $X_t$ and some candidate metrics, the graph refinement knowledge allows to filter out spurious relations, retaining only metrics that may have caused the anomaly.

\subsection{Causal Graph Discovery}
\label{sec:cauDis}
In this section, we describe our strategy to learn causal relations between the metrics in $\mathbf{X}$. 
Given $\mathbf{X} = \{X_1, X_2,\dots,X_N\}$ containing $N$ observed numerical time series $X_i \in \mathbb{R}^T$ of the same length $T \in \mathbb{N}^+$, the goal of \emph{causal discovery} is to learn the causal relations between all $N$ time series in $\mathbf{X}$ and the \emph{time-lag} between cause and effect, modeling them in a \emph{causal graph}. In a causal graph $\mathbf{G} = (V,E)$, each vertex $v_i \in V$ represents an observed time series $X_i \in \mathbf{X}$ and each edge $e = (i,j,\tau) \in E$ from $v_i$ to $v_j$, with $v_i, v_j \in V$, represents a causal relation where time series $X_i$ causes an effect in $X_j$ with time-lag $\tau \in [0, \dots, T-1]$. Note that the causal graph $\mathbf{G}_{0}$ induced by $\tau = 0$, i.e., $\mathbf{G}_{0} = (V,E_{0})$ with $E_{0} = \{(u,v, 0) \in E\}$, is a \emph{directed acyclic graph} (DAG).
To construct the causal graph, constraint-based causal discovery algorithms consider two phases: skeleton discovery, which learns causal relations between pairs of metrics with a conditional independence test, and edge orientation, which determines the direction of the causal relations learned in the previous phase. However, due to conflicting orientation rules, they may return undirected causal relations when it is impossible to determine which of two causally related metrics causes the other.

Our causal discovery algorithm is based on the PCMCI+ algorithm~\cite{runge2020discovering} which is able to handle time series data and allows to find lagged causal relations up to a $\tau_{max} \in \mathbb{N}$ time-lag, with $\tau_{max}$ provided in input by the user. 
By executing our algorithm, we obtain a causal graph $\mathbf{G} = (V,E)$, where:
\begin{itemize}
\item each vertex $v_i \in V$ represents a time series $X_i \in \mathbf{X}$;
\item each edge $e = (i,j,\tau) \in E$ from $v_i$ to $v_j$, with $v_i, v_j \in V$, represents a causal relation where time series $X_i$ causes an effect in $X_j$ with time-lag $\tau \in [0, \dots, \tau_{max}]$;
\item each edge $e = (i,j) \in E$ between $v_i$ and $v_j$, with $v_i, v_j \in V$, represents a contemporaneous undirected causal relation between time series $X_i$ and $X_j$.
\end{itemize}
Note that the edges $(i, j)$ and $(j, i)$ represent the same causal relation and only contemporaneous relations can be undirected, as the orientation of lagged relations is determined by the time-lag (an event cannot cause something occurred before it).

In the following section, we illustrate how we enhance the causal graph learned in this phase considering partial graph knowledge and an entropy based orientation strategy. 

\subsection{Causal Graph Enhancement}
\label{sec:cgdw}
In this section, we explain how \alg\ enhances the causal graph learned in section~\ref{sec:cauDis} considering the partial graph knowledge and an entropy based orientation strategy. Let us remember that causal discovery methods provide guarantees to recover the correct causal graph only under data assumptions~\cite{guo2020survey,pearl1995causal}, i.e., causal sufficiency, faithfulness, etc., which may be difficult to satisfy and verify in real-world scenarios. Thus, causal discovery algorithms may find spurious relations, miss true causal relations, and return undirected relations, as discussed in section~\ref{sec:cauDis}. 
However, by allowing partial domain knowledge as input, \alg\ enables domain experts to contribute to the causal graph. Thus, this phase aims to correct the potential spurious relations found by the causal discovery algorithm and to add expert-defined causal relations that the algorithm may have missed. Finally, an entropy based orientation strategy is used to orient any undirected edges.

\begin{algorithm}[]
	\KwData{Causal graph $\mathbf{G} = (V,E)$, domain directed edges $E_D$, domain node level function $\texttt{level}()$}
	\KwResult{Causal graph with domain knowledge $\mathbf{G_{DW}}$}
        $V_{DW} \leftarrow V$, $E_{DW} = \emptyset$\; \label{line:1}
	\ForEach{$(u,v) \in E_D$\label{line:2}}{
	   $E_{DW} \leftarrow E_{DW} \cup (u,v,0)$\; \label{line:3}
	}
        \ForEach{$(u,v,\tau) \in E$\label{line:4}}{ 
	   \lIf{$\texttt{level}(u) \leq \texttt{level}(v)$\label{line:5}}
        {
        $E_{DW} \leftarrow E_{DW} \cup (u,v,\tau)$}
	}
        \ForEach{$(u,v) \in E$\label{line:6}}{
	   \lIf{$\texttt{level}(u) < \texttt{level}(v)$\label{line:7}}
        {
        $E_{DW} \leftarrow E_{DW} \cup (u,v,0)$}
         \lIf{$\texttt{level}(u) > \texttt{level}(v)$\label{line:8}}
        {
        $E_{DW} \leftarrow E_{DW} \cup (v,u,0)$}
	}
        \small{$E_{DW} = \texttt{EntropyOrientation}(E_{DW}, E, \texttt{level}(\cdot))$\label{line:9}\;}
                \normalsize
	$\mathbf{G_{DW}} \leftarrow (V_{DW}, E_{DW})$\; \label{line:10}
	\Return $\mathbf{G_{DW}}$\;
	\caption{Causal graph enhancement.}
	\label{alg:enh}
\end{algorithm}

Given in input the causal graph $\mathbf{G} = (V,E)$ returned by the causal discovery algorithm and the partial graph knowledge, composed of the domain directed edges $E_D$ and the domain node level function $\texttt{level}(\cdot)$, in this phase \alg\ creates a \emph{causal graph with domain knowledge} $\mathbf{G_{DW}}=(V_{DW}, E_{DW})$ by enhancing the causal graph $\mathbf{G}$ with the partial graph knowledge.
The procedure to obtain $\mathbf{G_{DW}}$ is shown in Algorithm~\ref{alg:enh}. First, it sets $V_{DW} = V$ and $E_{DW} = \emptyset$ (line~\ref{line:1}). Then, it adds all instantaneous causal relations $(u,v) \in E_D$ defined by domain experts to $E_{DW}$ (lines~\ref{line:2}-\ref{line:3}). Thus, the final graph $\mathbf{G_{DW}}$ does not miss any true instantaneous causal relation defined by domain experts.
All directed (lagged) causal relations $(u,v,\tau) \in E$, with $\tau \in \{0, \dots, \tau_{max}\}$, that are not in contrast with the higher-order causal relations defined by the domain node level function (i.e., if $\texttt{level}(u) \leq \texttt{level}(v)$) are then added to $E_{DW}$ (lines~\ref{line:4}-\ref{line:5}), otherwise they are discarded since they represent spurious relations found by the algorithm. 
The domain node level function is also used to orient the contemporaneous undirected causal relations $(u,v) \in E$ when possible, i.e., when $\texttt{level}(u) \neq \texttt{level}(v)$ (lines~\ref{line:6}-\ref{line:8}).
Thus, by employing the partial domain knowledge, it is also possible to orient some causal relations that the causal discovery algorithm has not been able to orient.
Finally, for all the contemporaneous undirected causal relations $(u,v) \in E$ that cannot be oriented considering the domain node level function, i.e., if $\texttt{level}(u) = \texttt{level}(v)$, it employs the entropy based orientation strategy~\cite{kocaoglu2017entropic} (line~\ref{line:9}, see section~\ref{sec:entropy}). 
The final graph $\mathbf{G_{DW}}$ obtained with Algorithm~\ref{alg:enh} is the causal graph with domain knowledge. It represents all the causal relations defined in the domain knowledge and the ones learned from the causal discovery algorithm that are not in conflict with the partial graph knowledge, without unoriented relations.

\subsubsection{Entropy Based Orientation Strategy}
\label{sec:entropy}
The entropy based orientation strategy~\cite{kocaoglu2017entropic} is employed by \alg\ in the causal graph enhancement phase to orient undirected causal relations found in the causal discovery phase.
Given $u$ and $v$ two metrics connected by an undirected causal relation, the entropy based orientation strategy aims to compute the noise distribution of the two causal relations $(u,v,0)$ and $(v,u,0)$. The causal relation whose noise distribution has a lower entropy is considered correct. When applied to a pair of metrics, it returns the orientation of their relation and a score representing the level of confidence of such an orientation. If the score is too low, the strategy is considered inconclusive.

In the causal graph enhancement phase, \alg\ employs the entropy based orientation strategy to orient the causal relations $(u,v) \in E$ that cannot be oriented considering the domain node level function, i.e., if $\texttt{level}(u) = \texttt{level}(v)$, as follows.
For each contemporaneous undirected causal relation $(u,v) \in E$ such that $\texttt{level}(u) = \texttt{level}(v)$, it applies the entropy based orientation strategy, collecting the oriented causal relation and the score returned by the strategy. The causal relations with inconclusive orientations are discarded since the entropy based orientation strategy does not provide enough confidence about their orientations, and they may represent spurious relations found by the causal discovery algorithm. Then, it sorts all the oriented causal relations obtained in the previous step considering their score (higher scores first) to include the oriented causal relations in $\mathbf{G_{DW}}$ starting from those for which the entropy based orientation strategy is more confident. Thus, following the obtained ordering, it adds every oriented causal relation returned by the entropy based orientation strategy, i.e., $(u,v,0)$ or $(v,u,0)$, in $E_{DW}$ if the relation does not create any cycle in the sub-graph of $\mathbf{G_{DW}}$ induced by $\tau = 0$, i.e., in the sub-graph of $\mathbf{G_{DW}}$ with only contemporaneous causal relations. Oriented causal relations that create any cycle are instead discarded since they represent spurious relations found by the causal discovery algorithm.

\subsection{Causal Graph Refinement}
\label{sec:rcv}
In this section, we explain our strategy to find the initial set of candidate root causes $\mathcal{RC}$ associated with the anomaly of the performance metric. The idea is to find, between the candidate metrics, the ones that are more associated with the performance metric and finally to filter out the ones that do not respect the graph refinement knowledge.

For each metric $c \in \mathcal{C}$ in the candidate set, the idea is to compare it with the performance metric $X_t$ by computing their adjusted correlation score based on the Pearson correlation. The operation is repeated multiple times, first considering the original normalized data and then applying some data transformations, i.e., time shifting and smoothing. Finally, the maximum adjusted correlation score between those obtained with or without data transformations is considered for each candidate metric $c$.
The idea of data transformations (time shifting and smoothing) is to deal with noise and uncertainty in the data.
Time shifting aims to find time-lagged relations in the data. Since we are interested in finding causal effects of the candidate metrics on the performance metric $X_t$, we only consider negative shifts up to $maxShift$ steps to the candidate metrics, i.e., we compare earlier time samples of the candidate metrics to later time samples of the performance metric.
A \emph{shift penalty} based on $s \in \{0, \dots, maxShift\}$ is applied to the correlations found considering $s$ shift steps. 
Instead, the smoothing aims to correct for unwanted noise in the data. We apply smoothing to both candidate and performance metrics by considering a moving average window of size up to $maxWidth$ and only performance-candidate metric pairs with the same moving average window size are considered. 
A \emph{smoothing penalty} based on $w \in \{1, \dots, maxWidth\}$ is applied to the correlations found considering $w$ as moving average window size. (Note that $maxShift$, $maxWidth$, and the two penalties are configuration parameters.)

\begin{algorithm}[]
	\KwData{Performance metric $x$, candidate metric $c$}
	\KwResult{Triple $(score, corr, penalty)$ between $x$ and $c$}
 $scores \leftarrow \emptyset$\; \label{line2:1}
$x_n \leftarrow \texttt{normalize}(x)$, $c_n \leftarrow \texttt{normalize}(c)$\; \label{line2:2}
\For{$w \leftarrow 1$ \KwTo $maxWidth$ \label{line2:3}}{
            $x_{ns} \leftarrow \texttt{smooth}(x_n, w)$, $c_{ns} \leftarrow \texttt{smooth}(c_n, w)$\;\label{line2:4}
            $penalty_{smooth} = smoothPenalty \cdot (w-1)$\; \label{line2:5}
           \For{$s \leftarrow 0$ \KwTo $maxShift$ \label{line2:6}}{
            $c_{nst} \leftarrow \texttt{shift}(c_{ns}, -s)$\; \label{line2:7}
            $penalty = penalty_{smooth} + shiftPenalty \cdot s$\; \label{line2:8}
            $corr \leftarrow \texttt{correlation}(x_{ns}, c_{nst})$\;\label{line2:9}
            $score \leftarrow \texttt{abs}(corr) - penalty$\;\label{line2:10}
            $scores \leftarrow scores \cup (score, corr, penalty)$\;\label{line2:11}
            \normalsize
        }
        }
	\Return $(score, corr, penalty) \in scores$ with maximum score (for same maximum score, minimum penalty)\;
	\caption{Compute adjusted correlation score.}
	\label{alg:simScore}
\end{algorithm}

Algorithm~\ref{alg:simScore} shows the procedure employed by \alg\ to compute the adjusted correlation score between a candidate metric $c$ and the performance metric $X_t$. For a pair of performance-candidate time series, it first normalizes both by subtracting their mean and dividing by their standard deviation (lines~\ref{line2:2}). Then, it applies the smoothing data transformation to the two normalized time series, considering several moving average windows $w \in \{1, \dots, maxWidth\}$ (lines~\ref{line2:3}-\ref{line2:5}). Note that considering $w = 1$, we do not apply any smoothing, and thus we consider the original normalized data. After this phase, we get $maxWidth$ pairs (one pair of original normalized data and $maxWidth - 1$ pairs of smoothed data considering different window sizes). To each pair (smoothed or not), we then apply the time shifting data transformation (lines~\ref{line2:6}-\ref{line2:8}) to the candidate metric considering several negative time shifts $s \in \{0, \dots, maxShift\}$. Note that considering $s = 0$, we do not apply any time shift and we consider the original time order of the data. At this point, we get a total of $maxWidth \times (maxShift + 1)$ pairs, considering all possible combinations of the two data transformations and the original normalized data. For each pair, we then compute the correlation of the two (transformed) time series (line~\ref{line2:9}) considering the Pearson correlation.
The penalty and the correlation are then used to compute the adjusted correlation score by subtracting the penalty from the absolute value of the correlation (line~\ref{line2:10}). This score represents how similar the performance metric $X_t$ and the candidate metric $c$ are considering the applied data transformations (0 no similarity, 1 high similarity). The final adjusted correlation score between $X_t$ and $c$ is then the maximum score among the ones obtained considering all the $maxWidth \times (maxShift + 1)$ pairs (if more pairs have the same score, the pair with the lowest penalty is considered). 
Finally, the candidate metric $c$ is considered a candidate root cause associated with the anomaly of $X_t$ whether its adjusted correlation score is $\geq minSim$, with $minSim \in [0,1]$ the \emph{minimum similarity threshold} provided in input by the user, and whether the sign of the correlation associated with its adjusted correlation score respects the graph refinement knowledge described in section~\ref{sec:dcr}, if such domain knowledge is available. Repeating the procedure for all candidate metrics, we obtain the candidate root causes $\mathcal{RC}$.

\subsection{Causal Graph Subtraction}
In this section, we explain how \alg\ trims the causal graph with domain knowledge $\mathbf{G_{DW}}=(V_{DW}, E_{DW})$ considering the candidate root causes $\mathcal{RC}$.
The aim is to obtain the root cause causal sub-graph $\mathbf{G_{RC}}$ that represents the causal relations between the performance metric $X_t$ and the candidate root causes that have a causal path with it. In particular, we start processing the candidate root causes considering their distance to $X_t$, starting from the closest one. (Thus, considering their levels defined by the domain node level function, from higher to lower levels, and, for same level root causes, considering their adjusted correlation score with $X_t$, from higher to lower scores.) For each candidate root cause $r \in \mathcal{RC}$, we check whether there exists a causal path between $r$ and $X_t$. If there is only one causal path in $\mathbf{G_{DW}}$ between $r$ and $X_t$, we maintain such a path. If there exists more than one causal path between $r$ and $X_t$, we maintain the one that passes through more candidate metrics in $\mathcal{RC}$. If more than one causal path passes through the same number of candidate metrics in $\mathcal{RC}$, we maintain the shortest one. Considering the paths that pass through more candidate metrics, we allow \alg\ to consider the causal relations between as many candidate root causes as possible, to find how the anomaly has spread throughout the system. Instead, by considering the shortest path, we aim to include as few intermediate metrics as possible since they do not provide further information for the root cause analysis. Finally, if there is no causal path in $\mathbf{G_{DW}}$ between $r$ and $X_t$, $r$ is removed from the root causes $\mathcal{RC}$, since such a metric is not causally related to the performance metric.
The final root cause causal sub-graph $\mathbf{G_{RC}}$ is then the graph obtained by considering all the metrics, i.e., vertices, and all the causal relations, i.e., edges, visited connecting the candidate root causes and the performance metric with the strategy defined above, finally adding further causal relations $(u,v,\tau) \in E_{DW}$ whether both $u$ and $v$ are in $\mathbf{G_{RC}}$.

\section{Experimental Evaluation}
In this section, we report the results of our evaluation considering a system diagnostic scenario applied to a real advertising system. Since the amount of problematic cases and system settings in the real production system are quite limited, we also considered simulated data from ground-truth causal graphs of different sizes to represent different system complexity, to get a more comprehensive evaluation. 

\subsection{Baseline Algorithms, Parameters, and Environment}
We compared different versions of \alg\ with 4 baseline methods: CoFlux~\cite{su2019coflux}, $\varepsilon$-diagnosis~\cite{shan2019diagnosis}, the original PCMCI+~\cite{runge2020discovering}, and TCDF~\cite{nauta2019causal}. 
By comparing with CoFlux and $\varepsilon$-diagnosis, we aim to demonstrate that commonly used correlation based algorithms may detect many spurious root causes. Instead, \alg\ can provide better results considering the system causal structure.
Instead, by comparing with PCMCI+ and TCDF, we aim to demonstrate that our enhanced algorithm provides better results than general causal discovery based root cause analysis algorithms. Indeed, through the graph enhancement and refinement, and by leveraging causal domain knowledge, \alg\ is able to resolve spurious and undirected relations of standard causal discovery methods.

In all our experiments, for \alg, we used $\tau_{max} = 1$. $\tau_{max}$ is the maximum time-lag considered in the causal relations, and thus it must be defined based on previous knowledge about the data and the desired type of analysis. For example, $\tau_{max} = 0$ only considers instantaneous causal relations, while $\tau_{max} = 1$ also considers lagged causal relations between two consecutive time samples. Based on our knowledge about the real advertising system, the data is sampled every hour and it does not make sense to consider a time-lag $> 1$ since a change in a metric cannot be caused by something happened 2 hours early. Since both $maxShift$ and $maxWidth$ depend on the maximum time-lag, we set $maxShift = \tau_{max} = 1$ and $maxWidth = \tau_{max} + 1 = 2$, which results in a moving average window that considers the present and the previous time sample, i.e., 1 as maximum time-lag. Instead, we set $minSim = 0.5$ since this value represents the best trade-off between recall and precision, as shown in the next session. Finally, we set $shiftPenalty = 0.004$ and $smoothPenalty = 0.01$.
For CoFlux, we considered $maxShift = 1$ (in accordance with $\tau_{max} = 1$) and a minimum similarity threshold $= 0.5$, which provides the best results and is in accordance with $minSim = 0.5$. 
For PCMCI+, we considered the implementation provided in the Tigramite library~\cite{tigraRepo} using RobustParCorr as conditional independence test, $\tau_{max} = 1$, and letting PCMCI+ to automatically optimize the significance threshold used in the conditional independence test. (The same parameters have been considered also for PCMCI+ in the causal discovery phase of \alg.) Finally, for TCDF, we considered the implementation available online~\cite{TCDFRepo} with $kernel\_size = dilatation = 2$, which are the values suggested to consider 1 as maximum time-lag. All experiments have been executed on a machine with 16 GB of RAM and an Intel i7-11370H@3.30GHz using Python 3.10.

\subsection{Results with Simulated Data}
In this section, we report the results of our experimental evaluation with simulated data to assess the performance of \alg\ in detecting the right root causes considering different system complexities. The generated data is available online~\cite{OurRepo}.
The usage of simulated data is dictated by the need of having a ground-truth causal graph and an anomaly caused by a known root cause whose effect can be observed in a given performance metric. To generate simulated data, we considered four ground-truth causal graphs, respectively of 5, 10, 15, and 25 nodes, whose structures are based on the real advertising system of our use case scenario. From each of the four ground-truth causal graphs, we generated 50 time series datasets. Given a causal graph, we randomized the weights of its edges considering a uniform distribution between 0.1 and 0.9 to generate each of the 50 datasets. We then employed the data generator provided by the Tigramite library~\cite{tigraRepo} to generate a dataset composed of $T = 100$ samples in accordance with the causal graph. Finally, we injected an anomaly in a root cause metric picked at random among a set of possible root causes and propagated the anomaly to all its descendants up to the performance metric, considering the ground-truth causal graph and in accordance with the simulation process of the data. Note that this is a fair and realistic strategy to simulate anomalous data since an anomaly is an event that changes the distribution of the root cause. The anomaly then affects all the other components causally related to the root cause, i.e., all its descendants that have a causal path with it in the ground-truth causal graph. To respect the input requirement of \alg, the anomaly was injected only in the central portion of the dataset, obtaining $1/3$ of normal data before and after the anomaly, and $1/3$ of anomalous data.
For each of the four causal graphs, we then executed \alg\ and the baseline algorithms in all the $50$ datasets generated from it, and computed the average recall and precision obtained by each algorithm. The recall represents the ratio of runs in which the algorithm returned the correct root cause while the precision represents the average, over all the runs, of the ratio between the number of correct root causes and the number of total root causes returned by the algorithm. For PCMCI+ and TCDF, we considered as detected root causes all the components that had a directed causal path with the performance metric in the causal graph returned by the algorithm. Instead, for CoFlux, we considered all the components that had a directed edge ending in the performance metric or that had an undirected edge with it. (Note that we would have penalized CoFlux by not considering undirected edges, since it orients the edges only considering the time shift. In addition, since the edges are found considering the correlation, it did not make sense to consider causal paths of multiple edges.)
Finally, for $\varepsilon$-diagnosis, we directly considered the root causes reported by the algorithm. Since $\varepsilon$-diagnosis requires the same amount of normal and anomalous data, we executed it first considering the normal data before the anomaly and then the normal data after the anomaly, and reported the best results. (Note that in all the runs, the results of the two executions are very similar.) 

To evaluate the impact of the domain knowledge model, we compared different variations of \alg\ with different domain knowledge models in input: 
\begin{enumerate*}
\item \alg\ w/o DK: \alg\ without any domain knowledge;
\item $\alg (L)$: \alg\ with the domain node level function (higher-level causal relations between the metrics used to orient the edges, defined considering the respective ground-truth causal graph);
\item $\alg(L,kE)$: $\alg(L)$ with a random $k\%$ of the directed edges from the respective ground-truth causal graph included in the domain knowledge, with $k \in \{10,25,50\}$.
\end{enumerate*}
Since the four baseline methods do not allow domain knowledge, their results are obtained without any domain knowledge provided in input.
Finally, we considered an alternative version of \alg, called \alg\_P (and its variations \alg\_P$(L)$ and \alg\_P$(L,kE)$ with different domain knowledge models) in which we replaced the adjusted correlation score (see section~\ref{sec:rcv}) with the standard Pearson correlation computed between candidate and performance metrics, to show the benefits of the causal graph refinement.

\begin{figure}
        \centering
            \begin{subfigure}[b]{0.49\linewidth}
            \centering
            \includegraphics[width=\linewidth]{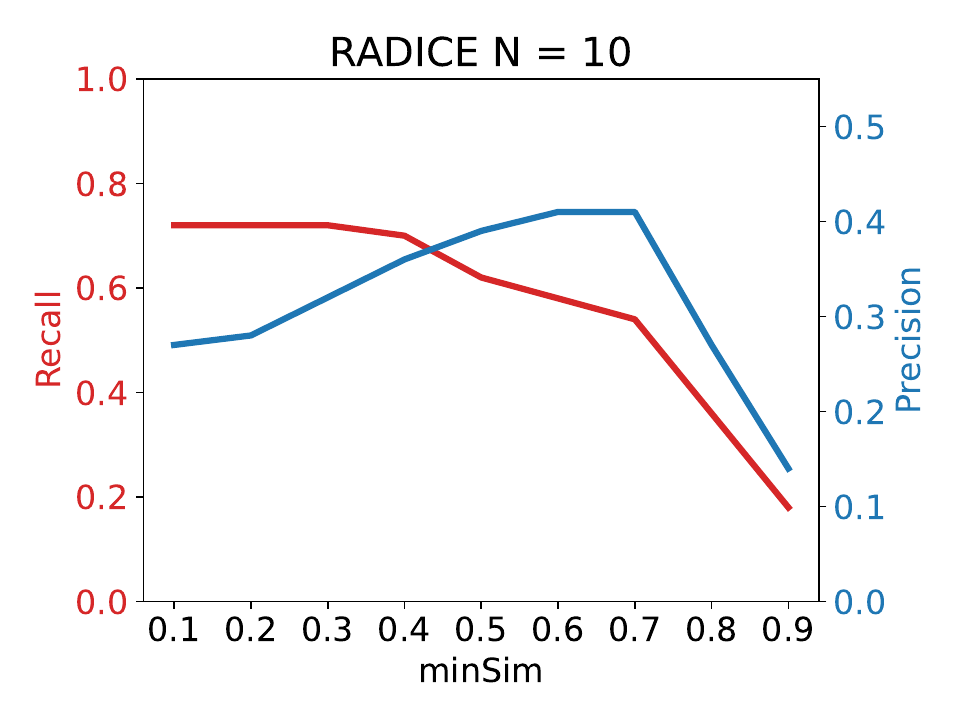}
    \end{subfigure}
        \hfill
        \begin{subfigure}[b]{0.49\linewidth}  
            \centering 
            \includegraphics[width=\linewidth]{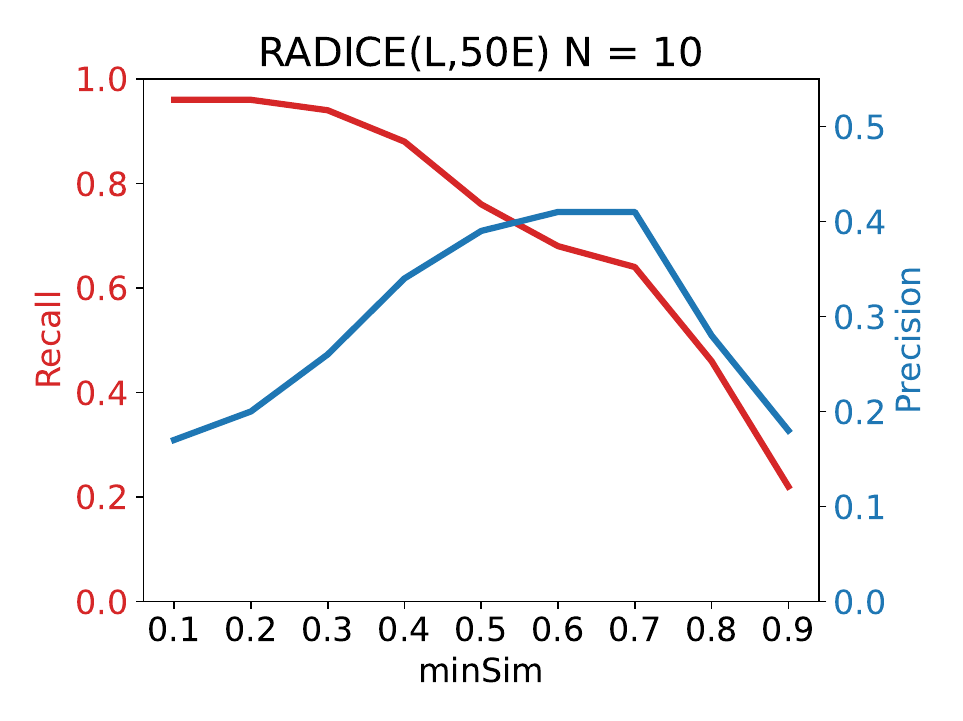}
    \end{subfigure}
   
        \begin{subfigure}[b]{0.49\linewidth}   
            \centering 
            \includegraphics[width=\linewidth]{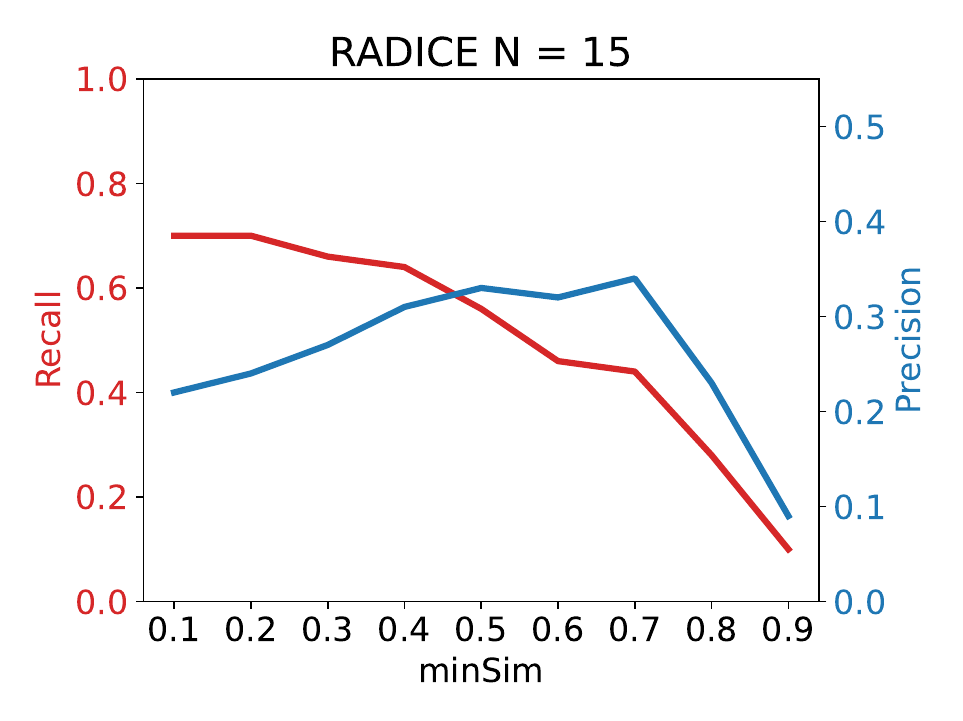}
\end{subfigure}
        \hfill
        \begin{subfigure}[b]{0.49\linewidth}   
            \centering 
            \includegraphics[width=\linewidth]{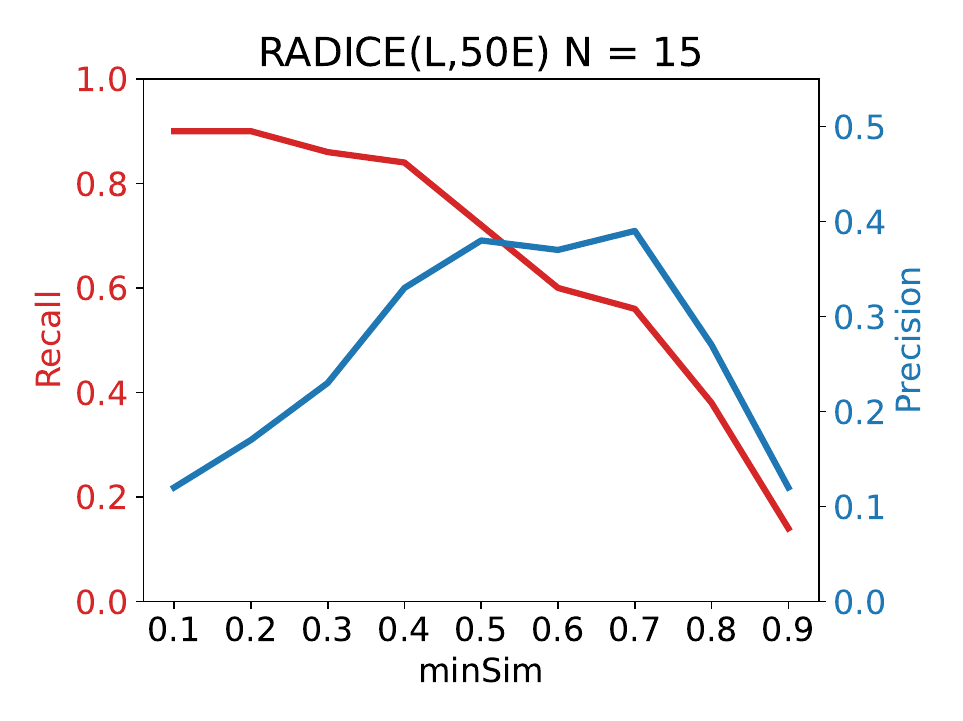}
\end{subfigure}
        \caption{Results for $minSim$. It shows recall and precision of \alg\ w/o DK and \alg$(L,50E)$ with simulated data (with $N=10$ and $N=15$ nodes) varying $minSim$.} 
        \label{fig:minSim}
    \end{figure}
    
First, we evaluated \alg\ varying the parameter $minSim \in \{0.1,0.2,\dots,0.9\}$ used in the graph refinement phase (see section~\ref{sec:rcv}). Figure~\ref{fig:minSim} reports the results of the simplest variation \alg\ w/o DK and of the variation with more domain knowledge $\alg(L,50E)$, for $N = 10$ and $15$. (The results of the other variations are analogous.) From the results, it is possible to notice that low values of $minSim$ guarantee higher recall but lower precision since only a few root causes are filtered out. By increasing the value of $minSim$, more metrics are filtered out and thus the recall decreases while the precision increases. Finally, for high values of $minSim$, both recall and precision decrease, since \alg\ reports none or a few root causes. Thus, we set $minSim = 0.5$ in the remaining of our evaluation since it provides the best trade-off between precision and recall.

Table~\ref{tab:res} reports the results of \alg\ and of the baseline algorithms in detecting the right root causes. From the results, it can be seen that \alg\ w/o DK always performs better than the original PCMCI+ algorithm, showing that the entropy based orientation strategy and the adjusted correlation score are enough to obtain improvements with respect to PCMCI+, even if no domain knowledge is available. \alg\ w/o DK performs also better than $\varepsilon$-diagnosis with all the ground-truth causal graphs. With $N = 10$ and $15$, it also performs better than CoFlux and TCDF, while, for $N = 25$, TCDF provides better recall but lower precision.
For $N = 5$, instead, the low effectiveness of PCMCI+ affects also \alg, which performs worse than TCDF, while CoFlux provides better recall but lower precision.
By including domain knowledge, the overall effectiveness of \alg\ improves, always outperforming the baseline algorithms.
The largest performance increase is provided by the node level function, with recall and precision increased on average by $50\%$ and $30\%$, respectively. By providing additional domain knowledge, i.e., directed edges, the recall almost always increases again, while the precision, in some cases, slightly decreases, since the added causal relations help in detecting the correct root cause, but may also implicate the addition of a few false positives. Furthermore, note that \alg\ (with and w/o DK) always performs better than its respective version \alg\_P, showing that the adjusted correlation score, which considers time shifting and smoothing, provides better results than the Pearson correlation. Finally, note that the precision is not very high overall. However, in a real-world scenario, the precision can be increased by considering the graph refinement knowledge that allows to remove further spurious root causes. (In these experiments, it is not possible to define refinement knowledge, since the weights of the edges are randomized making impossible to know in advance the correlation sign between metrics.)

\begin{table}[]
\caption{Results with simulated data. It shows average recall (R) and average precision (P) obtained with each algorithm for every ground-truth causal graph, with and without (w/o) domain knowledge (DK). In bold are highlighted the best results (with and w/o domain knowledge).}
\label{tab:res}
\centering
\resizebox{\linewidth}{!}{
\begin{tabular}{llcc|cc|cc|cc}
\toprule
                             &      & \multicolumn{2}{c}{$N = 5$}                                  & \multicolumn{2}{c}{$N = 10$}                                 & \multicolumn{2}{c}{$N = 15$}      & \multicolumn{2}{c}{$N = 25$}                                    \\ 
                                   \cmidrule(l{2pt}r{2pt}){3-4}\cmidrule(l{2pt}r{2pt}){5-6}\cmidrule(l{2pt}r{2pt}){7-8}\cmidrule(l{2pt}r{2pt}){9-10}
                            &       
                            & \multicolumn{1}{c}{R} & \multicolumn{1}{c}{P} %& \multicolumn{1}{c}{Ex (s)} 
                            & \multicolumn{1}{c}{R} & \multicolumn{1}{c}{P} %& \multicolumn{1}{c}{Ex (s)} 
                            & \multicolumn{1}{c}{R} & \multicolumn{1}{c}{P} %& \multicolumn{1}{c}{Ex (s)} 
                            & \multicolumn{1}{c}{R} & \multicolumn{1}{c}{P} %& \multicolumn{1}{c}{Ex (s)} 
                            \\ \midrule
\multirow{6}{*}{w/o DK}   
& CoFlux                             & \textbf{0.84}                      & 0.36        & 
\textbf{0.62}                      & 0.18        & 
0.48                      & 0.19     & 
0.15 &  0.10
\\
& $\varepsilon$-diagnosis                             & 0.64                      & 0.23        & 
0.38                      & 0.13        & 
0.26                     & 0.09     & 
0.14 &  0.05
\\
& PCMCI+                             & 0.12                      & 0.05           & 0.48                      & 0.14      & 0.40                      & 0.11      &  0.14 & 0.05     \\
& TCDF & 0.64       &  \textbf{0.48}  & 0.54  & 0.32  & 0.39  & 0.19   & \textbf{0.20} & 0.10 \\
 & \alg\_P w/o DK & 0.48                      & 0.45                & 0.60                      & 0.38                 & 0.46                      & 0.29           & 0.16 &  \textbf{0.11}     \\
& $\alg$ w/o DK & 0.58                      & 0.45         & \textbf{0.62}                      & \textbf{0.39}           & \textbf{0.56}                      & \textbf{0.33}        &  0.16 & \textbf{0.11}       \\

\midrule
\multirow{8}{*}{DK}
& \alg\_P$(L)$                         & 0.80                      & 0.48                 & 0.70                     & 0.40              & 0.54                     & 0.31           & 0.32 & 0.21      \\
& \alg\_P$(L,10E)$                   & 0.80                      & 0.48              & 0.70                      & 0.40            & 0.54                    & 0.31          &  0.36 & 0.22    \\
& \alg\_P$(L,25E)$                   & 0.82                 & 0.47               & 0.70                 & 0.40             & 0.56                   & 0.32         & 0.36 & 0.22    \\
& \alg\_P$(L,50E)$                   & 0.82                      & 0.47             & 0.72                      & 0.39              & 0.60                      & 0.34             &  0.38 & 0.22    \\ 
& $\alg (L)$                         & 0.82                      & \textbf{0.49}                & 0.74                      & \textbf{0.41}                 & 0.64                      & 0.35            & 0.36 & 0.22     \\
& $\alg (L,10E)$                   & 0.82                      & 0.48           & 0.74                      & \textbf{0.41}                & 0.64                      & 0.35       & 0.40 & 0.22       \\
& $\alg (L,25E)$                   & \textbf{0.86}                      & 0.48             & 0.74                      & \textbf{0.41}                 & 0.66                      & 0.35          & 0.40 & 0.22   \\
& $\alg (L,50E)$                   & \textbf{0.86}                      & 0.48              & \textbf{0.76}                      & 0.39             & \textbf{0.72}                      & \textbf{0.38}        & \textbf{0.42} & \textbf{0.23}       \\ \bottomrule
\end{tabular}
}
\end{table}

Table~\ref{tab:resTime} reports the execution times. For \alg\ only one time is reported, since those obtained with different variations are analogous. 
It is not surprising that \alg\ has the highest computational time since it employs both a causal discovery algorithm and a correlation based approach. However, it is still able to detect the root causes efficiently. The causal graph discovery and enhancement processes are the two steps with the highest computational complexity, while the computation of the adjusted correlation score has an execution time close to the one of the Pearson correlation. Indeed, \alg\ is only slightly slower than \alg\_P.

Overall, these results show that \alg\ is able to detect the right root causes in many cases and that its performance improves considering partial domain knowledge in input. Instead, causal discovery and correlation based methods alone lead to less accurate results. They also prove the validity of the adjusted correlation score and of our causal graph enhancement process, since even without any domain knowledge in input, \alg\ outperforms PCMCI+.
Finally, note that the performance of all methods drops conspicuously with the largest graph, highlighting that the development of causal based root cause analysis algorithms for larger systems is still an open challenge due to its high complexity. Adaptive pruning of causal graphs or hierarchical modeling could be valuable future directions for the diagnostic of larger systems.

\begin{table}[]
\caption{Computational time with simulated data. It shows the average execution time in seconds (s) obtained with each algorithm for every ground-truth causal graph.}
\label{tab:resTime}
\centering
\begin{tabular}{lcccc}
\toprule
            & \multicolumn{4}{c}{Execution Time (s)}  \\
                             &    $N = 5$                                 & $N = 10$                              & $N = 15$     & $N = 25$                                    
                            \\ \midrule

 CoFlux                             & 0.08                      & 0.15        & 0.22                      & 0.51     
\\
 $\varepsilon$-diagnosis                             & 0.15                      & 0.31        & 0.46                      & 0.77
\\
 PCMCI+                             & 0.56                      & 1.12           & 1.73                      & 4.52          \\
 TCDF & 1.02       &  2.35  & 3.41  & 8.31  \\
  \alg\_P& 2.51                      & 3.80                & 4.90                 & 7.81          \\
 $\alg$ & 2.73                      & 3.94         & 4.91                      & 7.86    
    \\ \bottomrule
\end{tabular}
\end{table}

\subsection{Use Case with Real Advertising Data}

\begin{figure}
\centering
\resizebox{\linewidth}{!}{
\begin{tabular}{cc|c}
\toprule
 & Case 1 & Case 2 
\\ \midrule

CoFlux
&
\minipage{0.5\textwidth}
\centering
  \includegraphics[width=\linewidth]{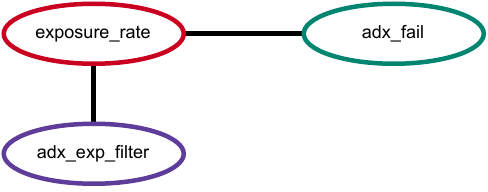}
  \endminipage
&
\minipage{0.5\textwidth}  
\centering
  \includegraphics[width=\linewidth]{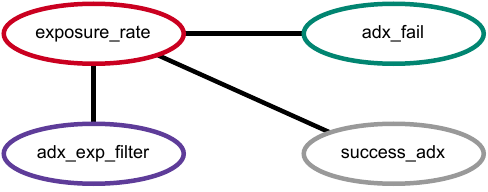}
\endminipage

\\ 
\midrule

PCMCI+
&
\minipage{0.5\textwidth} 
\centering
  \includegraphics[width=0.4\linewidth]{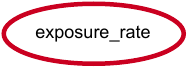}
  \endminipage
&
\minipage{0.5\textwidth} 
\centering
  \includegraphics[width=\linewidth]{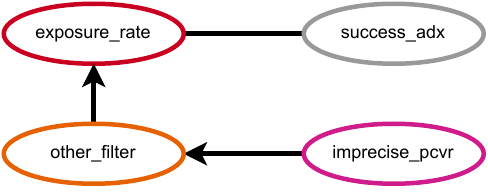}
\endminipage

\\ 
\midrule

TCDF
&
\minipage{0.5\textwidth}
\centering
  \includegraphics[width=\linewidth]{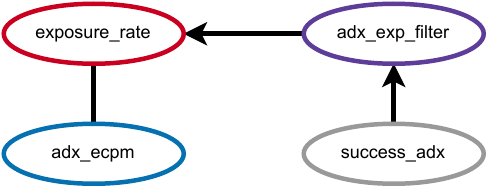}
  \endminipage
&
\minipage{0.5\textwidth}  
\centering
  \includegraphics[width=\linewidth]{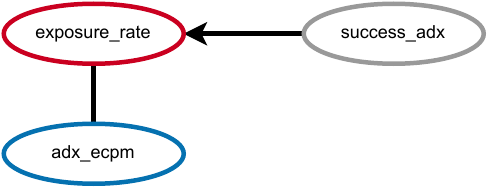}
\endminipage

\\ 
\midrule

\alg
&
\minipage{0.5\textwidth}
\centering
  \includegraphics[width=\linewidth]{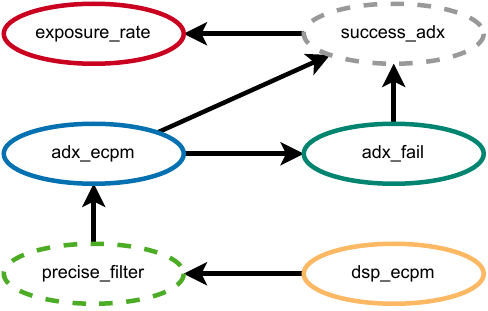}
  \endminipage
&
\minipage{0.5\textwidth}
\centering
  \includegraphics[width=\linewidth]{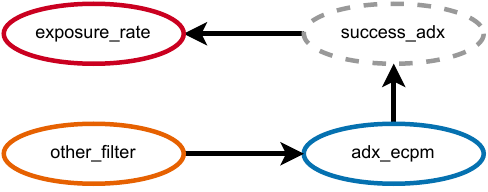}
\endminipage

\\ 
\bottomrule
\end{tabular}
}
\caption{Results with advertising data. It shows the causal sub-graph obtained with each algorithm for case 1 and case 2. Each metric is represented by a different color. 
For \alg, $exposure\_rate$ and the root causes are represented by continuous lines, while intermediate components by dashed lines. For CoFlux, PCMCI+, and TCDF, we only reported the portion connected with $exposure\_rate$.}
\label{fig:res}
\end{figure}

In this section, we share our experience considering a real advertising system to help readers understand \alg\ input, output, and performance in a real business scenario. The inputs of \alg\ are a time series for each metric in the system and the causal domain knowledge model that domain experts helped us to define, as described below. The output is instead the root cause causal sub-graph, containing performance, intermediate, and root cause metrics (shown in Fig.~\ref{fig:res}).
Our real data is composed of 3 days' data from a real advertising system (with a time sampling of 1 hour), in which the system exhibited two different anomalies in two different time windows. Note that the time sampling of 1 hour is dictated by the real system architecture which is independent of \alg. The data was initially composed of 19 metrics. 
In the data transformations, we converted all the metrics with integer values into rates considering their respective total counts, limiting their range between 0 and 1.
We considered as performance metric $exposure\_rate$ that represents the ratio of advertising messages that reached the users. Thus, an anomalous drop in this metric represents an issue in the overall system performance.
The remaining metrics represent different metrics of AD exchange (ADX) and demand-side platform (DSP) components of the advertising system: filtering rates, fail rates, success rates, effective cost per mille (ecpm), predicted conversion rate (pcvr), and predicted click-through rate (pctr).
The DSP is the proxy of advertisers, whose role is to bid on behalf of advertisers based on the estimated value of the impression and on environment insights, while the ADX connects advertisers and publishers serving as a trading desk that holds auctions~\cite{ou2023survey}.
From the transformed data, we then extracted the two anomalous periods associated with the two anomalies present in the data, i.e., the two periods in which $exposure\_rate$ had a drop. Both anomaly windows had a length of 8 hours, and thus of 9 time samples. Following the strategy described in section~\ref{sec:method}, we then created two datasets, each associated with an anomaly and composed of 27 time samples (9 samples before and after the anomaly, and 9 anomalous samples).
Having discussed with domain experts their knowledge about the system, we defined the partial domain knowledge, including 9 domain directed edges. The domain node level function divided the candidate metrics into 2 levels based on the type of components, ADX or DSP, with the latter that cannot be caused by the first ones. Instead, the graph refinement knowledge included a rule for each type of metric. For example, fail metrics (e.g., $adx\_fail$) with a positive correlation with $exposure\_rate$ cannot be root causes, since a drop in their values represents a drop in the fail rate of some components of the system that should increase the system performance and not cause a drop in $exposure\_rate$.

Figure~\ref{fig:res} reports the root cause causal sub-graphs obtained with \alg, CoFlux, PCMCI+, and TCDF with both anomalous datasets. (Since $\varepsilon$-diagnosis does not construct a root cause causal sub-graph, it has not been included in this analysis.) For both cases, \alg\ discovered (at least) a causal path between some candidate root causes and $exposure\_rate$. For case 1, \alg\ detected $dsp\_ecpm$, $adx\_ecpm$, and $adx\_fail$ as root causes. The graph refinement knowledge, instead, allowed to filter out $adx\_exp\_filter$, $success\_adx$, and $adx\_other\_filter$, all metrics with a high adjusted correlation score with $exposure\_rate$ (greater than $0.75$).
For case 2, \alg\ detected $adx\_ecpm$ and $other\_filter$ as root causes. The graph refinement knowledge, instead, filtered out two metrics that had an adjusted correlation score higher than those of the root causes, proving that high correlation does not imply causation. By showing \alg's results to domain experts, they confirmed the validity of our findings, since the reported causal relations and the detected root causes corresponded to those they found by manually analyzing the data.

CoFlux, for case 1, found two undirected links with $exposure\_rate$: while $adx\_fail$ is correctly indicated as a root cause, $adx\_exp\_filter$ represents a false positive (\alg\ filtered it out with the graph refinement knowledge for its positive correlation).
For case 2, instead, all three metrics connected with $exposure\_rate$ are false positives. In particular, \alg\ filtered out $success\_adx$ and $adx\_exp\_filter$ considering the graph refinement knowledge, while $adx\_fail$ did not have a sufficiently high adjusted correlation score to be considered.
PCMCI+ only in case 2 was able to find a directed causal path ending in $exposure\_rate$, correctly reporting $other\_filter$ as root cause, but wrongly including $imprecise\_pcvr$ and missing $adx\_ecpm$. (\alg\ filtered out $imprecise\_pcvr$ with the graph refinement knowledge.)
In addition, PCMCI+ detected many undirected causal relations (${\sim} 47\%$ of the reported relations) and found some spurious relations in contrast to domain knowledge (${\sim}12\%$ of the reported relations).
Finally, for case 1, TCDF wrongly reported $adx\_exp\_filter$ and $success\_adx$ as root causes. $Success\_adx$ is also the only root cause reported for case 2, but again represents a false positive. Note that TCDF found an undirected causal link with $adx\_ecpm$ in both cases, and the usage of a strategy such as the enhanced causal discovery we developed would allow to correctly flag it as a root cause.

Overall, these results highlight that \alg\ is able to report the correct root cause causal sub-graph in real industrial cases, since through the causal graph enhancement and refinement, and by leveraging domain knowledge, it can resolve many spurious and undirected relations.

\section{Conclusion}
In this work, we considered the problem of root cause analysis for system performance diagnostic. First, we introduced a domain knowledge model to allow domain experts to contribute partial causal knowledge for root cause analysis. Then, we proposed \alg, a causal graph based algorithm for root cause analysis for system performance diagnostic. 
\alg\ can employ (partial) domain knowledge as input and through the causal graph discovery, enhancement, refinement, and subtraction processes produces a root cause causal sub-graph showing the causal relations between metrics related to the anomaly. 
Our experimental evaluation with simulated and real data showed that \alg\ outperforms other baseline methods and that it is able to report root cause causal sub-graphs in real system performance diagnostic scenarios. 
Interesting future directions are the extension of \alg\ to directly consider domain knowledge in the causal discovery phase, the inclusion of probabilistic domain knowledge representing beliefs about causal relations between system components, and the development of algorithms whose performance does not drop conspicuously increasing the number of metrics, which is essential to diagnose larger complex systems.

\balance

\bibliographystyle{splncs04}
\bibliography{bib}

\end{document}